\documentclass{SCGE}

\emergencystretch = \maxdimen
\begin{document}

\begin{picture}(0,0){\rm
\put(0,-39){\makebox[160truemm][l]{\bf {\sanhao\raisebox{2pt}{.}}
Article {\sanhao\raisebox{1.5pt}{.}}}}}
\put(0,-52){\jiuwuhao {\textcolor[rgb]{0.5,0.5,0.5}{\sf 
}}}  
\end{picture}

\def\bm{\boldsymbol}

\def\dl{\displaystyle}
\def\du{\end{document}}
\def\pi{{\uppi}}

\Year{2012} %
\Month{??} %
\Vol{??} 
\No{??} 
\BeginPage{1} 
\EndPage{??} 
\AuthorMark{{\rm Gong X L, et al}}
\DOI{??} 

\title{Jet magnetically accelerated from disk-corona around
    a rotating black hole$^\dagger$}

\author[1*]{GONG XiaoLong}{}
\author[2]{LI LiXin}{}

\address[{\rm1}]{Department of Astronomy, Beijing Normal University, Beijing 100875, China;}
\address[{\rm2}]{Kavli Institute for Astronomy and Astrophysics, Peking University, Beijing
100875, P. R. China}

\maketitle \vspace{-3.5mm}{\footnotesize\begin{center} Received ?? ??, 2012; accepted ?? ??, 2012
\end{center}}\vspace*{-5mm}

\begin{center}
\rule{16.5cm}{0.4pt}
\parbox{16.5cm}
{\begin{abstract} A jet acceleration model for extracting energy from disk-corona surrounding
a rotating black hole is proposed. In the disk-corona scenario, we obtain the ratio of the power dissipated in the corona to the total for
such disk-corona system by solving the disk dynamics equations. The analytical expression of the jet power
is derived based on the electronic circuit theory of the magnetosphere. It is shown that jet power increases with the increasing black hole (BH)
spin, and concentrates in the inner region of the disk-corona. In addition, we use a sample consisting of 37 radio loud quasars to explore their jet production mechanism, and show that our jet formation mechanism can simulate almost all sources with high power jet, that fail to be explained by the Blandford-Znajek (BZ) process.
\end{abstract}}
\end{center}\vspace*{-0.6cm}

\begin{center}
\parbox{16.5cm}
{\bf\jiuhao Accretion and accretion disks, jets, corona
}
\end{center}

\begin{center}
{}
\vspace*{-0.1cm}
\Cit{Gong X L, Li L X. Jet magnetically accelerated from disk-corona around
    a rotating black hole. Sci China-Phys Mech Astron, 2012, ??: ??, doi: ??}
\end{center}
\wuhao\vspace*{1.5mm}

\begin{multicols}{2}

\renewcommand{\baselinestretch}{1.08} \baselineskip 12.2pt\parindent=10.8pt

\renewcommand{\thefootnote}


\sec{1\quad Introduction}

\no Jets exist in many astronomical cases, such as quasars, active galactic
nuclei (AGNs), and stellar binaries. Although the
precise mechanism for the acceleration and collimation of jets are still
unclear, the association of jets with magnetized accretion disks or
magnetized central objects is strongly supported by
the observations of Hubble Space Telescope, Chandra, and VLBI.
Some author [1-3] have agreed that jet formation should involve a large-scale magnetic field
threading an accretion disk or black hole (BH).

Several theoretical models [3-5] have been proposed for the acceleration and collimation of jets. These models
belong to two main regimes, the Poynting regime and the hydrodynamic regime. In the Poynting regime, energy is extracted in the form
of purely electromagnetic energy, but it is in the form of magnetically driven material wind in the

\vspace*{1mm}
\noindent\rule{2.5cm}{0.4pt}\\[0.1mm]{\qihao *Corresponding author (email:
gongxiaolong@mail.bnu.edu.cn)\vspace*{-1mm}\\
$\dagger$ Recommended by ZHU ZongHong (Editorial Board Member)}

\no hydrodynamic regime. Major progress has been made in the hydrodynamic regime of jet formation, while observed jets with bulk Lorentz factor $\Gamma \sim 10 $ in quasars and AGNs suggest that these jets are
likely to be in the Poynting regime [6].

In the Blandford-Znajek (BZ) process [4], energy and angular momentum can be extracted from a rotating BH
by an ordered poloidal magnetic field threading the BH horizon. Macdonald and Thorne [7] reformulated and extended the BZ
theory in a 3+1 split of the Kerr spacetime, and derived the analytic expressions of BH power and disk power by
using an equivalent electric circuit. The Blandford-Payne process [5] is also an important
mechanism for disk wind and jet production. In the Blandford-Payne process, energy and angular momentum are removed magnetically from accretion disk by the magnetic field lines that leave the disk surface and extend to large distances.

The relative importance of these two process is discussed by
different authors [1,8]. Ghosh and Abramowicz [8] argued that the BZ power
was overestimated, and it is, in general, dominated by the electromagnetic
power output of the inner region of the disk, provided that the poloidal magnetic field threading the BH does not differ significantly in strength
from that threading the disk. This argument is strengthened by the realization that the currents that generate the field threading the horizon must be situated in the disc rather than in the BH.  Li [9] reinvestigated the BZ process and discussed the magnetic coupling between the disk
and a rotating black hole based this equivalent electric circuit theory [10-12]. He proposed
that the toroidal electric current residing in a thin disk can generates a poloidal magnetic field threading the BH and disk [9]. The rotation of the
disk and BH can induce an electromotive force (EMF) on the disk and the BH's horizon. This EMF could be the energy source for the jets in AGN.
The results suggested that the BZ process may be less efficient in extracting energy from a rotating BH for the thin disk.

The large-scale magnetic field plays an important role in jet modelling.
Observations reveal that large-scale magnetic fields exist in compact
objects. The popular treatment of large-scale magnetic field is invoked
magnetohydrodynamical simulations. Though the origin of large-scale magnetic
field is still under controversy, some previous work [13,14] have proposed that the large-scale field can be
produced from the small-scale field created by dynamo processes. The length
scale of the fields created by dynamo processes is of the order of the disk
thickness $H$, and the poloidal component of the magnetic field is given
approximately by $B_P \sim (H / r)B_{dynamo} $. If the field is created in
the thin accretion disks ($H \ll r$), the large-scale field is very weak.
In the case of advection dominated accretion flow (ADAF) [15,16], the disk thickness $H \sim r$ and the poloidal component
of the magnetic field is stronger. In the disk-corona scenario, the
energetically dominant corona is the ideal situation for launching the powerful
jets/outflows [17]. The large-scale magnetic fields
created by dynamo processes in the corona are significantly stronger than
the thin disk due to the fact of the corona thickness being much larger than
the cold thin disk. Thus the corona can power a stronger jet than the thin
disk.

Recently Gong et al. [18] investigated a disc-corona model, in which part of
gravitational energy is dissipated in the hot corona, and simulated the emerged spectra from the disc-corona system for the different parameters using the Monte-Carlo method.

Motivated by the above works, in this paper we propose a coronal jet
model, in which a geometrically thin, optically thick disk surrounds a
rotating Kerr BH, and the corona is assumed to be heated by the reconnection
of the magnetic fields generated by buoyancy instability [19] in the disk. In our model, the large scale poloidal magnetic field can be derived from the magnetic energy density in the corona. We then get the coronal jet power using the equivalent electric circuit method in the magnetosphere. Finally, we use our jet formation mechanism to stimulate a sample consisting of 37 radio loud quasars. Throughout this paper the geometric units $G = c = 1$ are used.

\sec{2\quad Disk-corona model}

\noindent
In our accretion disk-corona system, a geometrically thin and optically
thick disk is sandwiched by a slab magnetic corona, and part of
gravitational energy of accretion matter is released in the hot corona. The
general relativistic model for a steady, axisymmetric, and thin Keplerian
disk around a Kerr black hole, has been investigated in detail by Novikov and Thorne [20]. In their model, it has been
assumed that there is no stress at disk's inner edge, that is, the so-called
``no-torque inner boundary condition''. In this case, the gas that reaches the stable circular orbit of minimum radius $ r = r_{ms}$, will ``fall out'' of the disk and spiral rapidly down the BH. Consequently, the gas density at $ r < r_{ms}$ is virtually zero compared to that at $ r > r_{ms}$, which means that no viscous stresses can act cross the surface $ r = r_{ms}$.

\vspace*{3mm}
The total gravitational power dissipated in unit surface area of the
accretion disk-corona system $Q$ is given by [21]

\begin{equation}
\label{eq1}
Q = (\dot {M} / 2\pi )e^{ - (\nu + \psi + \mu )}f,
\end{equation}

\noindent where $\dot {M}$ is the accretion rate of the disk, $\upsilon $, $\psi
$, $\mu $ are the metric coefficients, and the function of radius $f$ is defined as
\begin{equation}
\label{eq2}
f = - \frac{d\Omega }{dr}(E^ + - \Omega L^ + )^{ - 2} \int_{r_{ms} }^r {(E^+ } - \Omega L^ + )\frac{dL^ + }{dr}dr.
\end{equation}

\noindent The specific energy $E^{+ }$ and the specific angular momentum $L^{+}$ of a particle in the disk, can be written as

\begin{equation}
\label{eq3} E^\dag = {\left( {1 - 2\chi ^{ - 2} + a_ * \chi ^{ -
3}} \right)} \mathord{\left/ {\vphantom {{\left( {1 - 2\chi ^{ -
2} + a_ * \chi ^{ - 3}} \right)} {\left( {1 - 3\chi ^{ - 2} + 2a_
* \chi ^{ - 3}} \right)^{1 / 2}}}} \right.
\kern-\nulldelimiterspace} {\left( {1 - 3\chi ^{ - 2} + 2a_ * \chi
^{ - 3}} \right)^{1 / 2}},
\end{equation}

\begin{equation}
\label{eq4} L^\dag = M\chi {\left( {1 - 2a_ * \chi ^{ - 3} + a_ *
^2 \chi ^{ - 4}} \right)} \mathord{\left/ {\vphantom {{\left( {1 -
2a_ * \chi ^{ - 3} + a_ * ^2 \chi ^{ - 4}} \right)} {\left( {1 -
3\chi ^{ - 2} + 2a_ * \chi ^{ - 3}} \right)^{1 / 2}}}} \right.
\kern-\nulldelimiterspace} {\left( {1 - 3\chi ^{ - 2} + 2a_ * \chi
^{ - 3}} \right)^{1 / 2}},
\end{equation}

\noindent where $\chi = \sqrt {r / M} $ is the dimensionless radial coordinate, and
$a_\ast = a / M$ is the dimensionless black hole spin parameter.

\vspace*{2mm}
The power dissipated in the corona [17,18] is

\begin{equation}
\label{eq5}
Q_{cor} = P_m \upsilon _P = \frac{B^2}{8\pi }\upsilon _P,
\end{equation}

\noindent where $P_m $ is the magnetic pressure, and $\upsilon _P $ is the
velocity of magnetic flux transported vertically in the disk. Here we assume
the velocity $\upsilon _P $ of magnetic flux tubes is proportional to their
internal Alfven velocity, i.e. $\upsilon _P = b\upsilon _A $, and $b$ is
related to the efficiency of buoyant transport of magnetic structure.

\vspace*{3mm}
Now we give the equations of the disk structure as follows. The equation of
vertical pressure balance in the vertically-averaged form is

\begin{equation}
\label{eq6}
H = (P / \rho )^{1 / 2}(r^3 / M)^{1 / 2}AB^{ - 1}C^{1 / 2}D^{ - 1 / 2}E^{ -
1 / 2},
\end{equation}

\noindent where $H$ is the height of the accretion disk, $P$ and $\rho $ are pressure and
density of the disk respectively; $A$,$B$, $C$,$D$, $E$ are general
relativistic correction factors [20].

The equation of energy conservation is [20]

\begin{equation}
\label{eq7}
W = \frac{4}{3}(M / r^3)^{ - 1 / 2}CD^{ - 1}Q ,
\end{equation}

\noindent
where $W$ is integrated shear stress, defined as $W = 2\int_0^h {t_{r\varphi } } dz \sim 2t_{r\varphi } H$.
The interior viscous stress $t_{r\varphi } $  and the pressure $P$ are related by

\begin{equation}
\label{eq8}
t_{r\varphi} = \alpha P ,
\end{equation}

\noindent where $\alpha$ is the viscous parameter, and $\alpha = 0.3$ is adopted in our calculations.

The equation of state for gas on the disk is

\begin{equation}
\label{eq9}
P = P_{mag} + P_{tot} = P_{mag} + \frac{1}{3}aT^4 + \frac{\rho _0 T}{m_P },
\end{equation}

\noindent
where $P_{tot} $ is the total pressure (gas pressure plus radiation
pressure) at disk mid-plane, $a$ is the radiative constant.  $ m_p $ is the rest mass of proton, $ \rho _0$ and $ T$ are the density of rest mass
and the temperature in the disc, respectively. Here we assume

\begin{equation}
\label{eq10}
P_{mag} = \alpha_{0} \sqrt {P_{gas} P_{tot} }.
\end{equation}

The energy transport equation for the disk is

\begin{equation}
\label{eq11}
aT^4 = 2\kappa H\rho _0 (Q - Q_{cor} + \frac{1 - a_0 }{2}Q_{cor} )
\end{equation}

\noindent
where $a_0$ is the reflection albedo in the disk-corona. It is relatively low and most of the incident
photons from the corona are re-radiated as black body radiation [22]. $a_0 = 0.1$ is adopted in our model, and $\kappa $ is the Rosseland mean of total opacity [20].

\vspace*{2mm}

Solving eqs.(1)-(11) numerically, the power dissipated in the
corona $Q_{cor} $ and the structure of the disk can be derived as function of
radius $r$. The ratio of the power dissipated in the corona to the total for
such disk-corona system is given as

\begin{equation}
\label{eq12}
f_{cor} (r) = \frac{Q_{cor} (r)}{Q(r)} \quad .
\end{equation}

For the different black hole spin parameter $a_\ast $ , the curves of $ f_{cor} $ versus parameter $\xi $
defined as $\xi = r / r_{ms} $, and the accretion rate $\dot{m} = \dot{M} / \dot{M}_{Edd} $ are shown in Figure 1(a) and 1(b), respectively. It is shown that the ratio of the power dissipated in the corona $ f_{cor} $ increases with the increasing the spin parameter $a_\ast $, but decreases with the increasing accretion rate $\dot{m}$.


\begin{figure}[H]
\centering
{\includegraphics[scale=0.6]{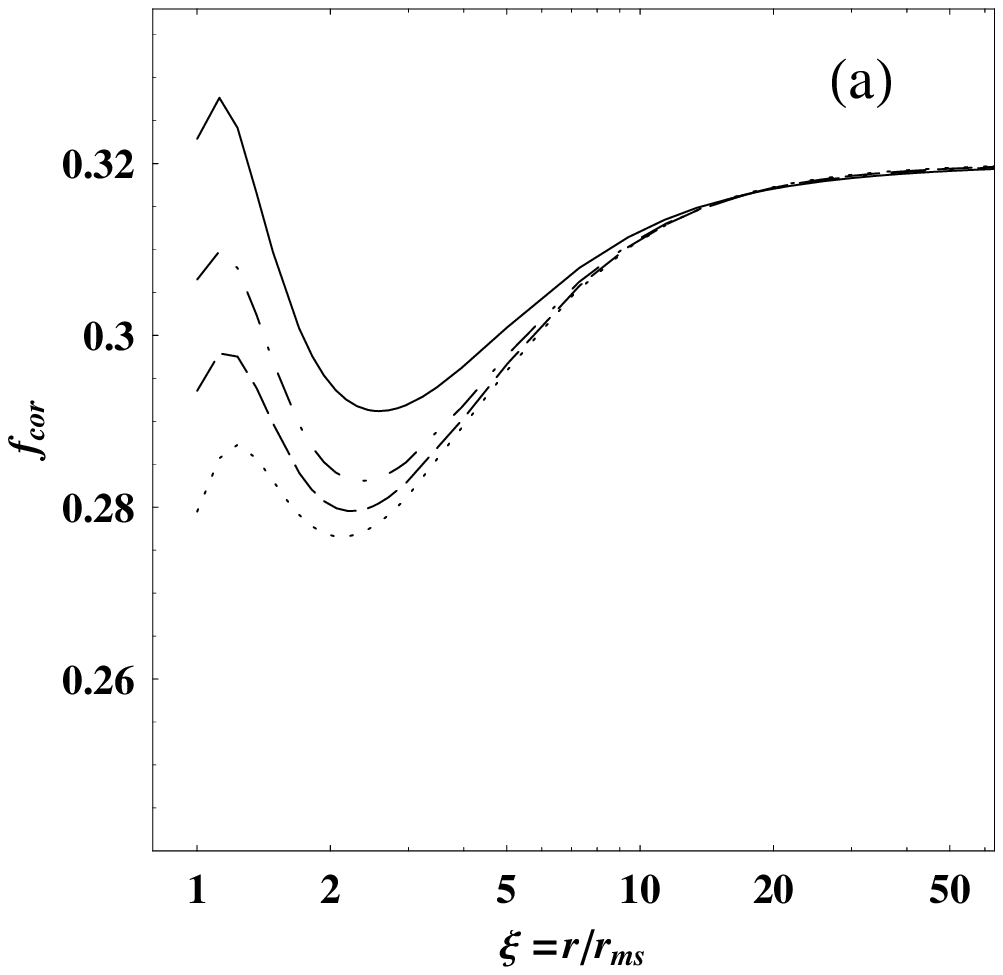}
 \includegraphics[scale=0.6]{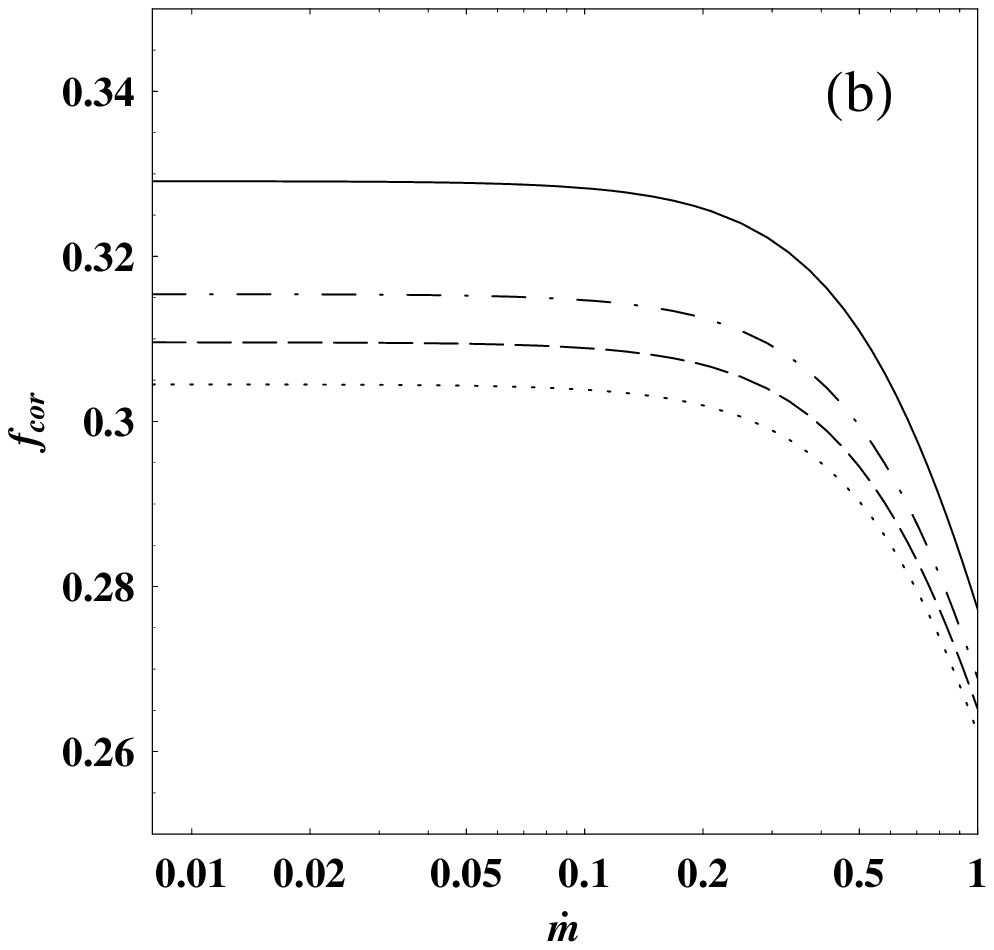}}
 \caption{(a) The value of $ f_{cor} $ varying with radial parameter $\xi $ for different values of $a_\ast $: $a_\ast = 0.1$ (dotted lines), $a_\ast = 0.5$ (dashed lines), $a_\ast = 0.7$ (dot-dashed lines), $a_\ast = 0.9$ (solid lines). $ \dot{m} = 0.1 $  is adopted in the calculations. (b) The value of $f_{cor} $ varies with $ \dot{m} $ for different values $a_\ast $. $ \xi = 30 $  is adopted in the calculations.}
\label{fig1}
\end{figure}

\sec{3\quad Jet from disk-corona}

\noindent
We wish to discuss the acceleration mechanism of the jets being magnetically
driven by the fields created in the corona. In this work, it is considered
that the coronal power $Q_{cor} $ can be partly dissipated locally to heat
the corona and ultimately radiated away as hard X-ray radiation with power
$L = \eta Q_{cor} $. Therefore we
assume that the magnetic energy density in the corona is

\begin{equation}
\label{eq13}
\frac{B_{cor}^2 }{8\pi } = \frac{ L(r)2\pi rdr}{2\pi rH_{cor} dr}t_0 ,
\end{equation}

\noindent where $H_{cor} $ is the height of the corona, and $t_0 \sim H_{cor} / v_{diss}
$ is the dissipation time. The dissipation velocity $v_{diss} $ depends on the
uncertain nature of the heating process, $v_{diss} = 0.01c$ is adopted in
our calculation.

Though the origin of large-scale magnetic field is still unclear,
some authors have verified that the large-scale field can be produced from the small-scale
field created by dynamo processes [13,14]. In the disk-corona system the strength of the poloidal field component depends on the typical scaleheight of a coronal magnetic flux tube and on the capability of reconnection events. In this paper, for simplicity, we assume that the poloidal component of the large-scale magnetic field is given approximately by

\begin{equation}
\label{eq14}
B_{cor}^P \simeq (H_{a} / r) B_{cor},
\end{equation}

\noindent
where $ H_{a} $ is the typical scaleheight of a coronal magnetic flux tube . As the height of the coronal magnetic flux tube is much larger than the disk thickness, we adopt $ H_{a} = 2 r_{ms}$ in our calculation.


\begin{figure}[H]
\centering
\includegraphics[scale=0.6]{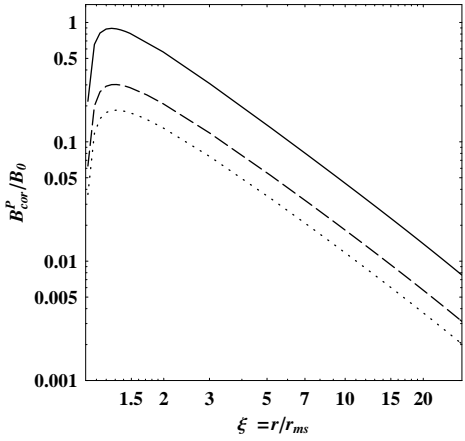}
\caption{The value of $ B^{P}_{cor} $ varying with radial parameter $\xi = r / r_{ms} $ for the different values of $a_\ast $ : $a_\ast = 0.1$ (dotted lines), $a_\ast = 0.5$ (dashed lines), $a_\ast = 0.9$ (solid lines). $\eta = 0.5$ is adopted in the calculations.}
\label{fig2}
\end{figure}

We plot the curves of the value of $ B^{P}_{cor} $ varying with radial parameter $\xi = r / r_{ms} $ for the different BH spin parameters $a_\ast $ in Figure 2. Note that the magnetic field $ B^{P}_{cor} $ is in units of $B_0 = \sqrt{\dot {M} / 8 \pi M^2} $ , where $\dot {M} $ is the accretion rate, and $M $ is the BH mass. Inspecting the Figure 2, we find that the value of the large scale poloidal magnetic field reaches a maximum as $ r = 1.3 r_{ms}$. In addition, the value of $ B^{P}_{cor} $ increases with the increasing BH spin parameter $a_\ast $.

Macdonald and Thorne [7] pointed that the magnetosphere anchored in a black hole and in its accretion disc should transfer much of the rotational energy of the hole and orbital energy of the disk into an intense flux of electromagnetic energy.  They constructed a general relativistic version of electronic circuit theory, and derived the analytical expressions of the electromagnetic power from the disk as follow [7]

\begin{equation}
\label{eq15}
\Delta P = I^2\Delta Z_A = \Omega _F (\frac{I}{2\pi })\Delta \psi .
\end{equation}

\noindent where $I$ is the equivalent current, $\Delta \Psi $ is the magnetic
flux between the two adjacent magnetic surfaces, and the quantity $\Omega _F $ is the magnetic field line angular velocity.

Based on the work of Macdonald and Thorne, we also propose an equivalent circuit to
calculate the electromagnetic power from the corona. The following equations
are used in deriving jet power,

\begin{equation}
\label{eq16}
\Delta Q_{jet} = \left( {I^p} \right)^2\Delta Z_{A} ,
I^p = {\Delta \varepsilon } \mathord{\left/ {\vphantom {{\Delta \varepsilon
} {\Delta Z_A }}} \right. \kern-\nulldelimiterspace} {\Delta Z_A },
\Delta \varepsilon = \left( {{\Delta \Psi } \mathord{\left/ {\vphantom
{{\Delta \Psi } {2\pi }}} \right. \kern-\nulldelimiterspace} {2\pi }}
\right)\Omega _F ,
\end{equation}

\noindent
where $I^p$ is the polodial current in each loop in the corona, $\Delta\varepsilon$ is the EMF due to the rotation of the disk-corona system. $\Delta \Psi $ and $\Delta Z_{A}$ are the magnetic flux between the two adjacent magnetic surfaces and the impendence of the corresponding acceleration-region, respectively. The magnetic flux $\Delta \Psi $ can be expressed as

\begin{equation}
\label{eq17}
\Delta \Psi = B_{cor}^p 2\pi \left( {{\varpi \rho } \mathord{\left/
{\vphantom {{\varpi \rho } {\sqrt \Delta }}} \right.
\kern-\nulldelimiterspace} {\sqrt \Delta }} \right)_{\theta = \pi
\mathord{\left/ {\vphantom {\pi 2}} \right. \kern-\nulldelimiterspace} 2}
dr.
\end{equation}

The magnetic field line angular velocity $\Omega _F $ is determined by the ratio of acceleration-region
impedance to disk impedance, and if the disk impedance is very low, $\Omega
_F \simeq \Omega _D $ [7]. In our disk-corona model, considering that corona impedance is also very low relative to the acceleration-region impedance, we can assume that the magnetic field line angular velocity in
the corona is
\begin{equation}
\label{eq18}
\Omega _F \simeq \Omega _D = \frac{1}{M(\xi ^{3 \mathord{\left/ {\vphantom
{3 2}} \right. \kern-\nulldelimiterspace} 2}\chi _{ms}^3 + a_ * )},
\end{equation}

\noindent where the radial parameter $\chi_{ms}$ is defined as $ \chi_{ms} = \sqrt{r_{ms} / M}$.
The load resistance $\Delta Z_A $ between the two adjacent magnetic surfaces
can be written as
\begin{equation}
\label{eq19}
\Delta Z_{A} \sim(5/6)(\Delta r / \varpi ) \sim(25\mbox{ }ohm)(\Delta r / \varpi ).
\end{equation}

The concerned Kerr metric coefficients are given as follows [23],
\begin{equation}
\label{eq20}
\left\{ {\begin{array}{l}
 \varpi = \left( {\Sigma \mathord{\left/ {\vphantom {\Sigma \rho }} \right.
\kern-\nulldelimiterspace} \rho } \right)\sin \theta , \\
 \Sigma ^2 \equiv \left( {r^2 + a^2} \right)^2 - a^2\Delta \sin ^2\theta ,
\\
 \rho ^2 \equiv r^2 + a^2\cos ^2\theta , \\
 \Delta \equiv r^2 + a^2 - 2Mr. \\
 \end{array}} \right.
\end{equation}

\noindent incorporating eqs. (16)--(20), and integrating eq.(16) over the
radial parameter $\xi = r / r_{ms} $, we have the expression for the jet power
as follows

\begin{equation}
\label{eq21}
Q_{jet} = \int_1^\xi {\frac{6 (B_{cor}^p M)^2(\xi^{3} \chi^{8}_{ms} + \xi a^{2}_{\ast} \chi^{4}_{ms} + 2 \chi^{2}_{ms} a^{2}_{\ast})^{3/2}  }{5 \xi^{1/2}(\xi^{2}\chi^{4}_{ms} - a^{2}_{\ast} - 2 \xi \chi^{2}_{ms})(\xi^{3/2} \chi^{2}_{ms} + a_{\ast}) }d\xi}.
\end{equation}
The power arising from the BZ process by a rotating black hole with a magnetic field $B_{\bot}$ normal to the horizon has been given by

\begin{equation}
\label{eq22}
P_{BZ} = \frac{1}{32} B^2_{\bot} \omega^{2}_{F} r^{2}_{H} a^{2}_{\ast},
\end{equation}

\noindent
where $ r_H $ is the horizon radius, and the factor $\omega^{2}_{F} \equiv \Omega_{F}(\Omega_{H} - \Omega_{F}) / \Omega^{2}_{H}$ is a measure of the effects of the angular velocity $\Omega_{F}$ of the field lines relative to that of the hole $\Omega_{H} $. In the case of $ \Omega_{F} = 1/2 \Omega_{H}$, the output power of BZ process takes a maximum, we adopt $ \omega_{F} = 1/2$.
Considering the balance between the magnetic pressure on the horizon and the ram pressure of the innermost part of an accretion
flow, Moderski et al. [24] give the relation between the magnetic field $B_{\bot}$ and the accretion rate $\dot {M} $ as:

\begin{equation}
\label{23}
(B_{\bot })^2 / 8\pi  = P_{ram} \sim \rho c^2 \sim \dot {M}/( 4\pi r^{2}_{H}).
\end{equation}
From eq.(23) the strength of magnetic field on BH horizon is
given by

\begin{equation}
\label{24}
B_{\bot} =\sqrt{2 \dot{M}/(M^{2} (1 + q)^{2})}, \quad q = \sqrt{1 - a^{2}_{*}}.
\end{equation}


\begin{figure}[H]
\centering
{\includegraphics[scale=0.6]{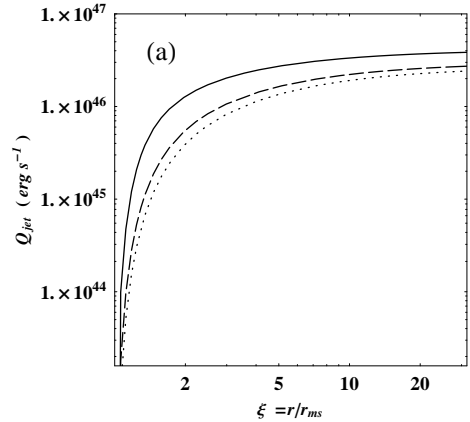}
 \includegraphics[scale=0.62]{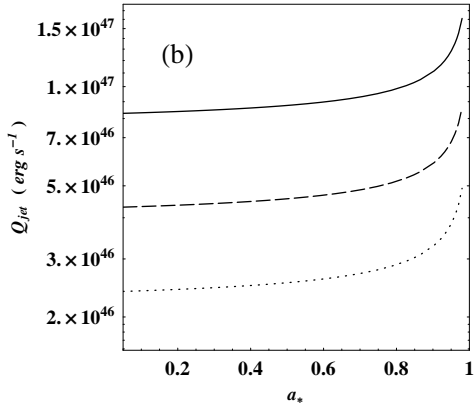}}
 \caption{(a): The curves of $ Q_{jet} $ versus radial parameter $\xi $ for different values of $a_\ast $: $a_\ast = 0.1$ (dotted lines), $a_\ast = 0.7$ (dashed lines), $a_\ast = 0.95$ (solid lines), and the accretion rate $\dot{m} = 0.1$ is assumed. (b): The curves of $ Q_{jet} $ versus $ a_\ast $ for different values of $ \dot{m} $: $\dot{m} = 0.1$ (dotted lines), $ \dot{m} = 0.2$ (dashed lines), $\dot{m} = 0.5$ (solid lines), and the radial parameter $\xi= 30 $ is assumed. $\eta = 0.5$, $ M = 10^{8}M_{\odot}$ are adopted in the calculations.}\label{fig3}
\end{figure}

The curves of the jet powers varying with the radial parameter $\xi
$ and the spin parameter $a$ are shown in Figure 3(a) and Figure 3(b). From the Figure 3, we
find that the jet power $Q_{jet} $ increases with the increasing black hole
spin $a_\ast $, the curves of $Q_{jet} $ turn to flat at $\xi \le 20$, imply
that the mostly jet power $Q_{jet} $ concentrates in the inner region of the disk. In
addition, the jet powers $Q_{jet} $ versus the accretion rate $\dot {m} = \dot{M} / \dot{M}_{Edd}$ for
the deferent parameters are plotted in Figure 4. It is evident that the jet
power increases monotonically with the increasing accretion rate.

\sec{4\quad Comparison with observation}

\noindent
The method of estimating jet power is important for quantifying the power emerging from the central engine
of the radio source. Willott et al. [25] used the optically thin flux density from lobes measured at 151 MHZ to estimate the value of the jet power, they have given an empirical relation between the jet power and the extended radio luminosity $ L_{151}$.  Recently Punsly [26] presented a theoretical derivation of an estimate for a radio source jet kinetic luminosity, which assumes that most of the energy in the lobes is in plasma thermal energy
with a negligible contribution from magnetic energy. Based on the formula derived by Punsly [26], Liu et al. [27] estimated the jet powers of 146 radio-loud quasars.
As the central black hole masses are given for the sources, the accretion rates of these sources can be estimated by using the bolometric luminosity
$L_{bol}$. The dimensionless accretion rate is given by
\begin{equation}
\label{25}
\dot{m} = \dot{M} / \dot{M}_{Edd} \simeq L_{bol} / L_{Edd}.
\end{equation}


\begin{figure}[H]
\centering
\includegraphics[scale=0.62]{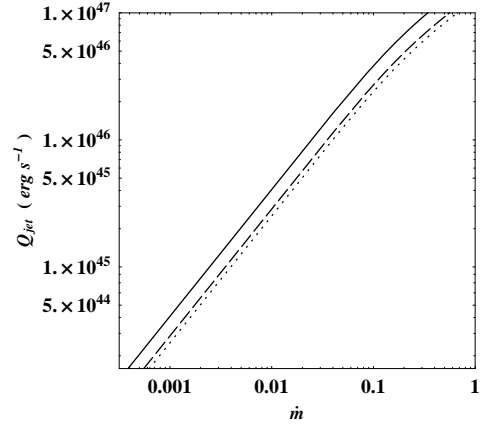}
\caption{The curves of $ Q_{jet} $ versus accretion rate $ \dot{m} $ for the different values of $a_\ast $ : $a_\ast = 0.1$ (dotted lines), $a_\ast = 0.7$ (dashed lines), $a_\ast = 0.95$ (solid lines). $\eta = 0.5$, $ M = 10^{8}M_{\odot} $, and $\xi= 30 $ are adopted in the calculations.}
\label{fig4}
\end{figure}

To investigate the jet acceleration mechanism, we compile a sample consisting of 37 radio loud quasars with high jet power.
The sample includes 14 steep-spectrum sources and 23 flat- spectrum sources. We adopt the estimated values of the BH mass given in ref.[28]. The estimated jet powers are taken from ref.[27]. The bolometric luminosities are taken from Woo and Urry [29], they are estimated via $ L_{bol} = 9 L_{5100}$ [30]. Our sample is shown in Table 1.

In Figure 5 and 6, we plot the relation between $L_{bol} / L_{Edd}$ and jet
power $Q_{jet} $. The jet power can be extracted by the large scale magnetic
field is calculated by using the eq.(21). The power extracted from a rapidly
rotating BH (i.e. the BZ power) is calculated by using the eq.(22). The
steep-spectrum sources and flat-spectrum sources in Figure 5 and 6 are labeled as
circles and triangles, respectively.

Inspecting the above Figures, we have the following results: (I) The jet in the great majority of sources cannot
be powered by the BZ process, even if the black hole spin parameter is $a_{\ast} =
0.99$; (II) The jet production mechanism in our model be able to produce
sufficient power observed in the 35 sources ( except 0538 + 498,1828 + 487 ), if black hole
spin parameter is taken into account in the calculation. In addition, we
find the jet power in our model depends mainly on the black hole spin $a_{\ast}$
and the accretion rate $\dot {m} \simeq L_{bol} / L_{Edd} $. By using
eq.(21), we stimulate the jet powers observed in these sources as shown in
the column (8) of Table 1. The reason that the two sources ( 0538 + 498,1828 + 487 ) cannot be fitted by our model, may be these two sources have different accretion modes, for example, ADAF or adiabatic inflow-outflow solutions [31]. Furthermore, we can not anticipate that the jet powers of all quasars are explained by one kind of jet acceleration mechanism.

\begin{figure}[H]
\centering
\includegraphics[scale=0.60]{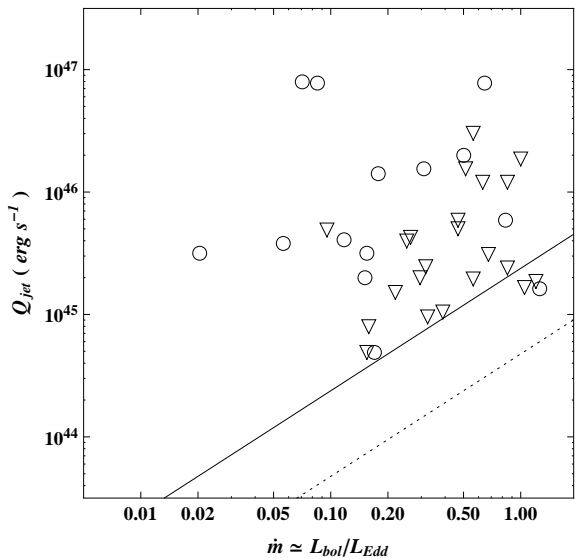}
 \caption{The relation between $L_{bol} / L_{Edd} $ and jet
power$Q_{jet} $. The solid line ($ M = 5\times 10^{9}M_{\odot} $) and the dotted line ($ M = 10^{8}M_{\odot} $) represent the jet powers produced by BZ process for $a_\ast = 0.99 $. }\label{fig5}
\end{figure}

\sec{5\quad Discussion}

\noindent
In this paper a jet model for extracting energy from disk-corona surrounding
a rotating BH is discussed. The expression for jet power $Q_{jet} $
is derived based on an electronic circuit in theory of the magnetosphere. Furthermore,we use a sample of 37 radio-loud quasars to explore
their jet production mechanism, and find that our jet formation mechanism can simulate almost
sources with high power jet, that can not explained by the BZ power.

However some uncertainties exist in our model. One uncertainty is the
geometry of the hot corona on the disk. For simplicity, we adopt
a slab corona in this paper. Some authors also propose a patchy corona,
which made of a number of separate active regions [17,32]. In that patchy-corona
scenario, if the numbers of active regions between the two adjacent magnetic
surfaces locate on the disk radius $r$ and $r + dr$ can be given, the jet power
should be calculated by using the equivalent electronic circuit method in
principal. Another uncertainty lies in the parameter $\eta $. Since much about the corona in the disk-corona system remains unknown, for example, the
fraction of power heating the corona is unclear. To simplify, we adopt $\eta = 0.5$ in the calculations.

\begin{figure}[H]
\centering
\includegraphics[scale=0.60]{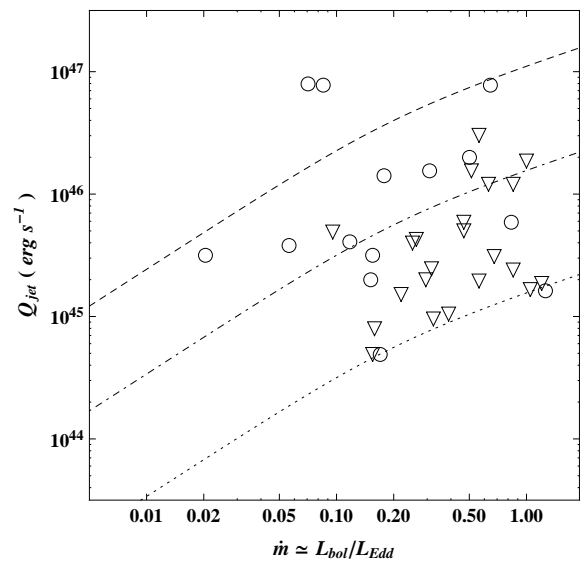}
 \caption{The relation between $L_{bol} / L_{Edd} $ and jet
power$Q_{jet} $. The dotted line ($ M = 10^{8}M_{\odot} $), the dot-dashed line ($ M = 10^{9}M_{\odot} $) and the dashed line ($ M = 5\times 10^{9}M_{\odot} $) represent the jet powers for  $a_\ast = 0.5 $ in our model.}\label{fig6}
\end{figure}

It should be noticed that the plunging region between the BH's horizon and the inner edge of the disk usually be neglected in
discussing the relative importance of BZ process by some author [1,8]. Li [33] proposed that the energy can be continuously extracted from the BH through this region and this may be the more efficient way for extracting energy from a Kerr BH. Reynolds et al. [34] argued that the plunging inflow can greatly enhance the trapping of large scale magnetic field on the black hole, and therefore may increase the importance of the BZ process relative to previous estimates that ignore the plunging region. Conversely, if some magnetic field lines connect the plunging region to the disk, this magnetic fields can exert stresses on the inner edge of an accretion disk around a BH [35]. This magnetic torque can considerably
enhance the amount of total energy released in the disk-corona system [36]. In such case, the magnetic energy density and the large scale magnetic field would be enhanced in our model. This would in turn increase the jet power from the corona. This effect of the plunging region on the BZ power and coronal jet power should be further investigated in the disk-corona scenario.

Finally, in this paper we focus mainly on the jet formation in the form of
Poynting fluxing. We shall discuss the acceleration of jet in the hydrodynamic regime
in future work.

\vspace*{2mm} \Acknowledgements{\bahao This work was supported by the National Basic Science Program (Project 973) of China (Grant Nos. 2009CB24901, 2012CB821804), the National Natural Science Foundation of China (Grant No. 10973003), and the Project for Excellent Young and Middle-Aged Talent of Education Bureau of Hubei Province (Grant No. Q2007120001).}

\normalsize \vskip0.3in\parskip=0mm \baselineskip 18pt
\renewcommand{\baselinestretch}{1.1}\footnotesize\parindent=4mm\bahao


\REF{1\ }Livio M, Ogilvie G I, Pringle J E. Extracting energy from black holes: the relative importance of the Blandford-Znajek mechanism. Astrophys J, 1999, 512: 100--104

\REF{2\ }Meier D L. A magnetically switched, rotating black hole model for the production of extragalactic radio jets and the Fanaroff and Riley class division. Astrophys J, 1999, 522: 753--766

\REF{3\ }Lovelace R V E. Dynamo model of double radio sources. Nature 1976 262: 649--652

\REF{4\ }Blandford R D, Znajek R L. Electromagnetic extraction of energy from Kerr black holes. Mon Not R Astron Soc, 1977, 179: 433--456

\REF{5\ }Blandford R D, Payne D G. Hydromagnetic flows from accretion discs and the production of radio jets. Mon Not R Astron Soc, 1982, 199: 883--903

\REF{6\ }Lovelace R V Eoldoba A V, Ustyugova G V et al. Poynting jets from accretion disks. Astrophys J, 2002,572: L445--L448

\REF{7\ }Macdonald D, Thorne K S. Black-hole electrodynamics - an absolute-space/universal-time formulation.
Mon Not R Astron Soc, 1982, 198: 345--382

\REF{8\ }Ghosh P, Abramowicz M A. Electromagnetic extraction of rotational energy from disc fed black holes - The strength of the Blandford-Znajek process. Mon Not R Astron Soc, 1997, 292: 887--895

\REF{9\ }Li L X. A toy model for Blandford-Znajek mechanism. Phys. Rev. D, 2000, 61: 0840161--0840167

\REF{10\ }Li L X, Pacy\'{n}ski B. Extracting energy from accretion into Kerr black hole.
 Astrophys J, 2000, 534: L197--L200

\REF{11\ }Li L X. Accretion Disk Torqued by a Black Hole. Astrophys J, 2002, 567: 463--476

\REF{12\ }Li L X. Observational signatures of the magnetic connection between a black hole and a disk. Astron Astrophys, 2002, 392: 469--472

\REF{13\ }Romanova M M, Ustyugova G V, Koldoba A V,et al. Dynamics of magnetic loops in the coronae of accretion disks. Astrophys J, 1998, 500:703--713

\REF{14\ }Tout C A, Pringle J E. Can a disk dynamo generate large-scale magnetic fields? Mon Not R Astron Soc, 1996 281: 219--255

\REF{15\ }Narayan R, Yi I. Advection-dominated accretion: Self-similarity and bipolar outflows.  Astrophys J, 1994, 444: 231--243

\REF{16\ }Narayan R, Yi I. Advection-dominated Accretion: Underfed Black Holes and Neutron Stars. Astrophys J, 1995, 452, 710--735

\REF{17\ }Merloni A, Fabian A C. Coronal outflow dominated accretion discs: a new possibility for low-luminosity black holes?
Mon Not R Astron Soc, 2002 332: 165--175

\REF{18\ }Gong X L, Li L X, Ma R Y. A disc-corona model for a rotating black hole. Mon Not R Astron Soc, 2012, 420: 1415--1422

\REF{19\ }Balbus S A, Hawley J F. Instability, turbulence, and enhanced transport in accretion disks. Revs. Mod. Phys.,1998, 70: 1--53

\REF{20\ }Novikov I D, Thorne K S. Astrophysics of black holes. In: Dewitt C, Dewitt B S, eds. Black Holes 1999(New York: Gordon and Breach)

\REF{21\ }Page D N, Thorne K S. Disk-accretion onto a black hole: time-averaged structure of accretion disk. Astrophys J, 1974 191, 499--506

\REF{22\ }Zdziarski A A, Lubinski P, Smith D A. Correlation between Compton reflection and X-ray slope in Seyferts and X-ray binaries.
 Mon Not R Astron Soc, 1999, 303: L11--L15

\REF{23\ } Thorne K S, Price R H, Macdonald D A. ed. Black Holes: The Membrane Paradigm, New Haven: Yale Univ. Press, 1986

\REF{24\ } Moderski R, Sikora M, Lasota J P. 1997, in Ostrowski M, Sikora M,
Madejski G, Belgelman M, eds, Proc. International Conf., Relativistic Jets in AGNs. Krakow, p.110

\REF{25\ } Willott C J,  Rawlings S, Blundell K M, et al. The emission line¨Cradio correlation for radio sources using the 7C Redshift Survey. Mon Not R Astron Soc, 1999, 309: 1017--1033

\REF{26\ } Punsly B.  An independent derivation of the Oxford jet kinetic luminosity formula. Astrophys J, 2005, 623: L9--L12

\REF{27\ } Liu Y, Jiang D R, Gu M F. The jet power, radio loudness, and black hole mass in radio-loud Active Galactic Nuclei
Astrophys J, 2006 637: 669--681

\REF{28\ } Gu M F, Cao X W, Jiang D R. On the masses of black holes in radio-loud quasars. Mon Not R Astron Soc, 2001, 327:1111--1115

\REF{29\ } Woo J H, Urry M C. AGN black hole masses and bolometric luminosities. Astrophys J,2002 579: 530--544

\REF{30\ } Kaspi S, Smith P S, Netzer H, et al. Reverberation measurements for 17 Quasars and the Size-Mass-Luminosity relations in Active Galactic Nuclei. Astrophys J, 2000, 533: 631--649

\REF{31\ } Blandford R D, Begelman M C, On the fate of gas accreting at a low rate on to a black hole
Mon Not R Astron Soc, 1999, 303: L1--L5

\REF{32\ }Merloni A, Fabian A C. Thunderclouds and accretion discs: a model for the spectral and temporal variability of Seyfert 1 galaxies. Mon Not R Astron Soc, 2001 328: 958--968

\REF{33\ }Li L X. Extracting energy from black hole through transition region.
 Astrophys J, 2000b, 540: L17--L20

\REF{34\ }Reynolds C S, Garofalo D, Begelman M C. Trapping of Magnetic Flux by the Plunge Region of a Black Hole Accretion Disk. Astrophys J, 2006, 651: 1023--1030

\REF{35\ } Agol E, Krolik J H. Magnetic stress at the marginally stable orbit: altered disk structure, radiation, and black hole spin evolution. Astrophys J,  2000, 528: 161--170

\REF{36\ } Gammie C F. Efficiency of Magnetized Thin Accretion Disks in the Kerr Metric.
Astrophys J,  1999, 522: L57--L60


\begin{table*}
\centering
\begin{minipage}{140mm}
\caption{Data of the sample and the values of the concerned parameter for fitting the jet power.}
\begin{tabular}{@{}llrrrrlrlr@{}}
  \hline\hline
   Sources & Type & $Log M_{BH}$ & $Log L_{bol}$ & $\dot{m}$ & $Log Q_{jet}$ & $a_{*}$ & $Log Q_{jet}(model)$\\
   (1) & (2) & (3) & (4) & (5) & (6) & (7) & (8)\\
 \hline
 0022-297 & SS & 7.91 & 44.98 & 0.12 &45.61 & 0.63 & 45.60\\
 0056-001 & FS & 8.71 & 46.54 &0.68 & 45.69 & 0.72 &45.70\\
 0119+041 & FS & 8.38 & 45.57 &0.15 &44.69 & 0.75 &44.69\\
 0133+207 & SS & 9.52 & 45.83 &0.02& 45.50 & 0.88&45.49  \\
 0134+329 & SS & 8.74 &46.44 & 0.50 &46.30 & 0.95 & 46.27  \\
 0135-247 & FS & 9.13 &46.64 &0.32& 44.98 & 0.18 & 45.02  \\
 0336-019 & FS & 8.98 & 46.32&0.21 & 45.18 & 0.17& 45.15  \\
 0403-132 & FS & 9.07  & 46.47 &0.25& 45.60  & 0.68&45.62 \\
 0405-123 & FS & 9.47 & 47.40 &0.85& 46.08 & 0.28&46.06  \\
 0518+165 & SS & 8.53 & 46.34 &0.65& 46.89 & 0.78& 46.90   \\
 0538+498 & SS & 9.58 & 46.43 &0.07& 46.90 & ---- & ----   \\
 0637-752 & FS & 9.41 & 47.16&0.56 & 46.48  & 0.75& 46.50  \\
 0838+133 & FS & 8.52 & 46.23 &0.51& 46.19  & 0.95&46.15  \\
 0923+392 & FS & 9.28 & 46.26 & 0.10 & 45.69 & 0.85&45.66 \\
 0954+556 & FS & 8.07 & 46.54 &3.00& 45.64 & 0.90&45.61  \\
 1007+417 & SS & 8.79 & 46.74 &0.83& 45.77 & 0.76 & 45.80 \\
 1100+772 & SS & 9.31 & 46.49 &0.15 &45.30 & 0.31& 45.35 \\
 1136-135 & FS & 8.78 & 46.78 &1.00& 46.27 & 0.91&46.28 \\
 1137+660 & SS & 9.36 & 46.85 & 0.31 & 46.19 & 0.85& 46.18 \\
 1250+568 & SS & 8.42 & 45.61 & 0.15 & 45.50 & 0.75& 45.50 \\
 1253-055 & FS & 8.43 & 46.10 &0.46 & 45.70  & 0.88&45.70  \\
 1354+195 & FS & 9.44 & 47.11 &0.47& 45.77 & 0.28&45.76 \\
 1355-416 & SS & 9.73 & 46.48 & 0.06 & 45.58 & 0.28&45.58 \\
 1611+343 & FS & 9.57 & 46.99 &0.26& 45.63  & 0.22&45.63 \\
 1637+574 & FS & 9.18 & 46.68& 0.31 & 45.39 & 0.20&45.38  \\
 1641+399 & FS & 9.42 & 46.89 &0.30& 45.30 & 0.18&45.32 \\
 1642+690 & FS & 7.76 & 45.78 &1.04& 45.22 & 0.84&45.21 \\
 1656+053 & FS & 9.62 & 47.21 &0.39& 45.02 & 0.21&45.07  \\
 1828+487 & SS & 9.85 & 46.78 &0.09& 46.89  & ----&----  \\
 2135-147 & SS & 8.94 & 46.17 &0.17& 44.69 & 0.30&44.70 \\
 2155-152 & FS & 7.59 & 45.67 &1.20 & 45.27 & 0.96 &45.24  \\
 2201+315 & FS & 8.87 & 46.62 &0.56 & 45.29  & 0.18&45.28 \\
 2216-038 & FS & 9.24 & 47.17 &0.85& 45.38  & 0.17&45.40 \\
 2251+158 & SS & 9.17 & 47.27 &1.26& 45.21  & 0.21&45.22  \\
 2255-282 & FS & 9.16 & 46.96&0.63 & 46.08  & 0.27 &46.06\\
 2311+469 & SS & 9.30 & 46.55 & 0.18 & 46.15 & 0.86& 46.18 \\
 2345-167 & FS & 8.72 & 45.92 & 0.16 & 44.90 & 0.62&44.89  \\
\hline
\end{tabular}

\medskip
\textbf{Notes:} Column(1):IAU source name. Column(2): the radio type (FS:flat-spectrum sources,SS: steep-spectrum sources ).Column(3): the black hole mass in the units of solar mass. Column(4): the bolometric luminosity of in the units of $ erg s^{-1}$ . Column(5): the accretion rate $\dot{m}$. Column(6): the jet power $Q_{jet}$ in unit of $erg s^{-1}$.  Column(7) is the value  of the BH spin parameter. Column(8): the fitted jet power.\\\\

\end{minipage}
\end{table*}

\end{multicols}

\end{document}